%% file: main.tex
\documentclass[10pt,twocolumn]{article}
\usepackage[margin=0.8in]{geometry}
\usepackage{amsmath,amssymb}
\usepackage{booktabs}
\usepackage{graphicx}
\usepackage{xcolor}
\usepackage{xurl}
\usepackage[hidelinks]{hyperref}
\usepackage{microtype}
\usepackage{tikz}
\usetikzlibrary{arrows.meta}

\makeatletter
\renewcommand\section{\@startsection{section}{1}{\z@}{-3.5ex \@plus -1ex \@minus -.2ex}{2.3ex \@plus .2ex}{\normalfont\Large\bfseries\raggedright}}
\renewcommand\subsection{\@startsection{subsection}{2}{\z@}{-3.25ex \@plus -1ex \@minus -.2ex}{1.5ex \@plus .2ex}{\normalfont\large\bfseries\raggedright}}
\makeatother

\title{Recompute or Restore? Host-Tier KV for Multi-Model Agent Workflows\\on a Static-Graph NPU Serving Stack}
\author{Sungwon Woo \quad Seyeon Kim\thanks{Korea University. Corresponding author: Seyeon Kim.}}
\date{}

\begin{document}
\maketitle

\begin{abstract}
Agent workflows issue dependent model calls that share long context prefixes, so serving them efficiently depends on reusing the key-value (KV) cache of those prefixes. On accelerators that compile model graphs ahead of time, however, the device KV budget is fixed, and prefixes evicted under load must be recomputed. Restoring them from host memory avoids this recomputation, but the transfer occupies the same engine thread as prefill and decode, so its benefit is not guaranteed. In this paper, we study when restoring beats recomputing on a commercial static-graph neural processing unit (NPU) stack (Rebellions ATOM+, vllm-rbln 0.10.2) that serves a Qwen3-8B planner on two cards and a Qwen3-0.6B executor on one. We first show that repeated prefill dominates planner work when default serving collapses: 89.5\% of the executed planner prompt tokens had been materialized before. We then build a synchronous host-DRAM tier on the runtime's public KV copy calls, together with a scheduler-aware event model whose predictions are frozen before each test cohort. Experimental results show that the host tier raises workflow service-level objective (SLO) attainment from 22.8\% to 100\% and from 9.4\% to 99.8\% in the two tested overload conditions, and endpoint confirmation on two new seeds supports conservative load gains of 11.8\% (chain) and 18.2\% (fork), while mixed and bursty workloads remain seed-sensitive. The frozen model predicts median and 99th-percentile (P99) latency within its 15\% tolerance in all 24 common-load conditions, including the 16--18\% P99 increase that the tier causes for fork workflows, but it misorders three policy contrasts and overestimates the benefit of a cheaper store path (11\% predicted against 3.1\% measured). The trade-off therefore depends on runtime scheduling and transfer cost as well as cache capacity.
\end{abstract}

\section{Introduction}

Agent workflows execute small programs of large language model (LLM) calls. A planner reads a task and a shared document, executors handle sub-steps, and an answer step reads the accumulated context again. Because consecutive calls share long prefixes, caching the key-value (KV) state of those prefixes~\cite{kwon2023vllm,zheng2024sglang} keeps such workloads affordable, and recent agent-serving systems schedule around it~\cite{lin2024parrot,luo2025autellix,pan2025kvflow,chen2026concur,kang2026thunderagent}.

Serving these workflows on an accelerator that compiles model graphs ahead of time raises a different problem. On such a neural processing unit (NPU), the model graphs and the KV budget are fixed at compile time, so the device cache has a fixed size and the vendor runtime owns the KV layout. When the prefixes of completed requests are evicted, later calls must recompute them. A host-memory tier that keeps those prefixes is a known remedy on GPUs~\cite{lmcache2025,hicache2025,vllmoffload}, and the NPU vendor documents a CPU offload path for a newer software stack~\cite{rblnoffload}.

\begin{figure*}[t]
\centering
\includegraphics[width=\textwidth]{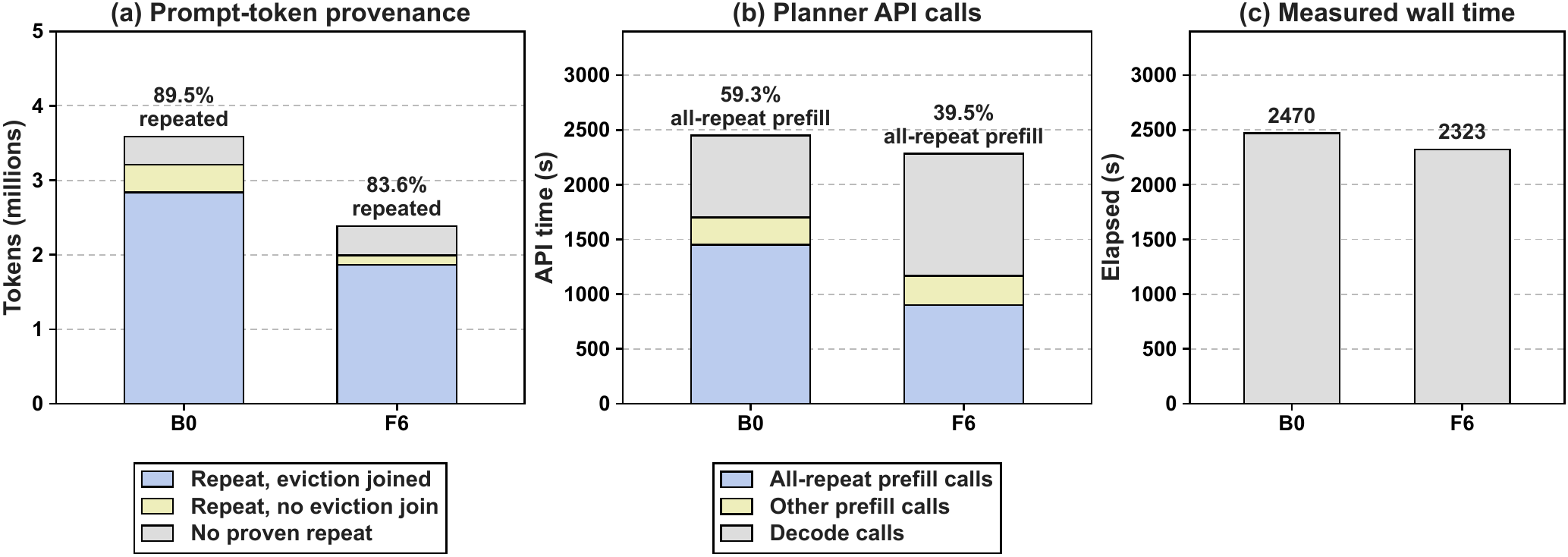}
\caption{Planner work in the KV64 fork collapse run (seed 501, 0.55 workflows/s). Under stock serving (B0), 89.5\% of the executed prompt tokens are repeated (a), and prefill calls made only of repeated tokens take 59.3\% of the host-inclusive API time (b). Repeated tokens are classified from earlier runtime completions and block history. The panels use different numerators and denominators, and API time includes warmup and failed-client requests, so it is not divided by the client wall time in (c).}
\label{fig:lineage}
\end{figure*}

However, two questions remain open before a host tier can be relied on in such a stack. First, it is unclear which work actually dominates when serving collapses under overload, and therefore how much a host tier can remove. Second, restoration has a cost of its own: storing a prefix takes time on the same engine thread that runs decode, and restoring it takes time before the next call can run. Whether restoring beats recomputing for a given workload is thus a quantitative question that depends on how the engine schedules prefill, decode, and transfers.

To answer these questions, we present a measurement study with prospective validation on one pinned serving stack, a Rebellions ATOM+ NPU server that runs a Qwen3-8B planner and a Qwen3-0.6B executor. The study rests on two components: (i) a host-tier KV bridge that connects vLLM's offloading manager to the runtime-owned KV layout and restores evicted prefixes from host DRAM, and (ii) a scheduler-aware discrete-event model that is calibrated on one seed and whose predictions are frozen before each test cohort. The paper does not propose a new cache or tiering design. Instead, it measures what restoration recovers, what it costs, and how far these effects can be predicted in advance.

Experimental results show that redundant prefill is the dominant work under overload and that restoring it is effective. In a collapsed run with default serving, 89.5\% of the executed planner prompt tokens had been materialized before, and the host tier raises workflow service-level objective (SLO) attainment from 22.8\% to 100\% and from 9.4\% to 99.8\% in two overload conditions. Endpoint confirmation on two new seeds supports conservative load gains of 11.8\% for chain and 18.2\% for fork workflows. At common load, the frozen model predicts every median (P50) and 99th-percentile (P99) latency within 15\%, including a 16--18\% P99 increase that the tier causes for fork workflows, although it misorders three policy contrasts and overestimates the benefit of a cheaper store path.

The main contributions of this paper are as follows:

\begin{itemize}
\item We identify the work that dominates an overload collapse on a static-graph NPU serving stack. A lineage decomposition that joins client requests to scheduler decisions, runtime calls, and block evictions shows that repeated prefill dominated the measured planner work in the studied collapse run, with no request preemption (\S\ref{sec:lineage}).
\item We build a host-restore adapter for the runtime-owned KV layout of this pinned stack, with each ordering rule marked as inherited from vLLM's offloading manager or adapted, and a scheduler-aware event model of its serving behavior (\S\ref{sec:design}).
\item We evaluate both in three frozen-prediction cohorts that cover common load, capacity endpoints, and an implementation change to the store cost. We report numerical errors, policy-ordering failures, and output differences separately, including the cases in which the model was wrong (\S\ref{sec:eval}).
\end{itemize}

\section{Background and Motivation}
\label{sec:background}

This section describes how a static-graph NPU serves agent workflows and then examines, on a collapsed run, which work a host tier would have to remove.

\subsection{Serving Agent Workflows on a Static-Graph NPU}
\label{sec:npu}

An agent workflow is a small directed acyclic graph (DAG) of dependent model calls that share long prompt prefixes. On the stack we study (\S\ref{sec:setup}), each model is compiled with a fixed KV budget. The default, which we call KV32, is the SDK's automatic value of 32K tokens per model, and a 64K planner budget (KV64) is used only in the first two cohorts of Table~\ref{tab:cohorts}. Decode graphs exist for batch sizes 1, 2, and 4, whereas prefill runs one request at a time in fixed 128-token chunks. A long uncached prompt therefore runs as consecutive 128-token calls without returning control to the scheduler. The runtime owns the KV layout and exposes it only through copy calls. Consequently, when the prefix of a completed request is evicted, a later call that needs it must recompute it.

\subsection{Redundant Prefill Under Overload}
\label{sec:lineage}

\begin{figure*}[t]
\centering
\input{figs/fig_overview.tex}
\caption{Overview of the serving path (top) and the prospective evaluation (bottom), with a chain workflow as the dependency example. On the planner, restore, prefill, store, and decode share one engine thread: the full host prefix is restored when the usable host prefix is longer than the local one ($H_h>H_l$), and missing host entries are stored after prefill. Boxes show the order of operations, not their durations. The event model makes predictions from the frozen trace and does not control serving.}
\label{fig:system-overview}
\end{figure*}
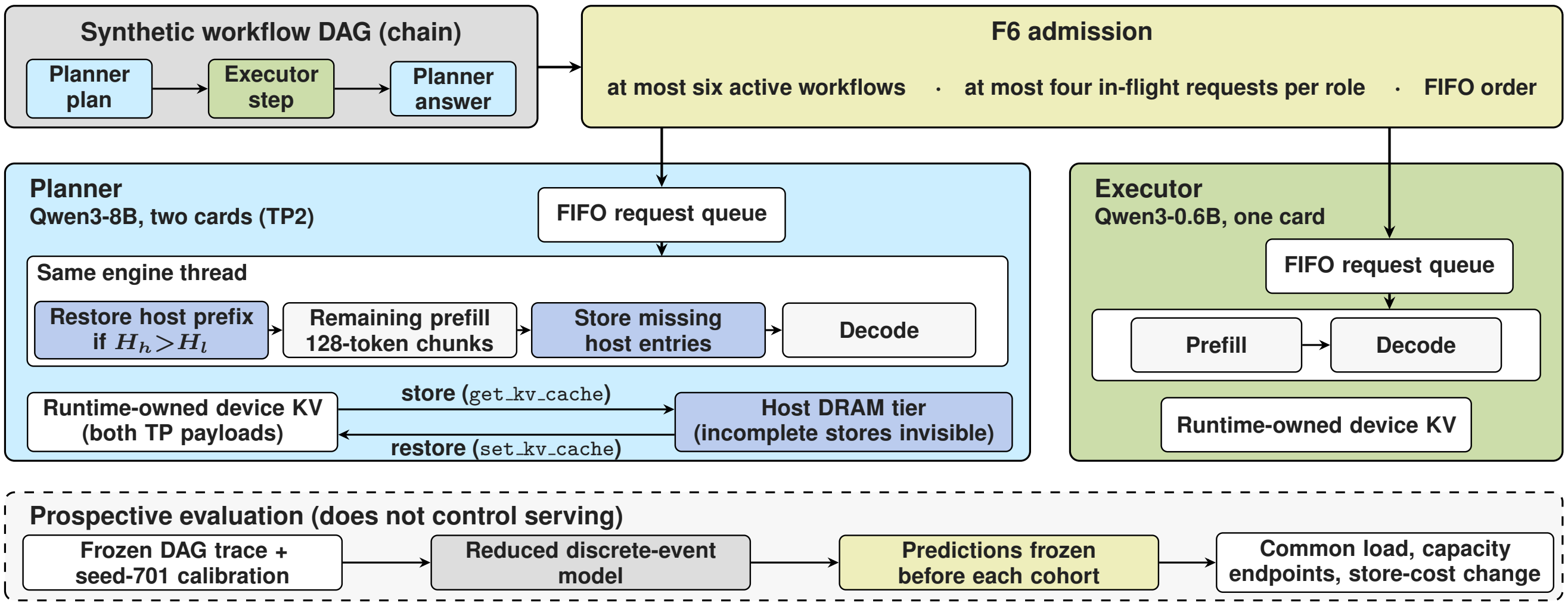

To determine what a host tier could recover, we first examine which work dominates when default serving collapses. We added an observer to the vLLM engine that records, for every request, its scheduling, the runtime calls it took part in, and the blocks allocated to and evicted from it. In two 48-workflow pairs with the observer on and off, no records were lost, outputs matched the uninstrumented server for 192 of 192 requests, and P99 rose by 1.4\% and 6.6\%; a later check closed the remaining gap between metadata assignment and runtime completion.

Figure~\ref{fig:lineage} shows the result for the KV64 fork workload under stock serving at 0.55 workflows/s, a run that collapsed (\S\ref{sec:setup} defines the workloads and policies). In this run, 89.5\% of the planner prompt tokens that were executed had been materialized before, and 79.2\% could be joined to an observed eviction. Prefill calls whose tokens were entirely repeated took 1,452.93 of 2,449.71\,s of summed host-inclusive planner runtime-call time (59.3\%). Moreover, no running request was preempted, so the repeated work came from prefixes of \emph{completed} requests that later requests rebuilt. Role times overlap in wall time and the token and API-time fractions have different denominators, so we do not add them, and the API-time shares do not translate directly into latency savings. The restore runs of \S\ref{sec:recovery} use the same configuration but are separate executions, so we do not pair them with this run. Nevertheless, this observation motivates keeping evicted prefixes in host memory instead of recomputing them.

\subsection{Cost of Restoring from Host Memory}
\label{sec:cost}

A host tier removes this recomputation, but it adds work of its own. On this stack, the copies that move KV between the device and host memory occupy the same engine thread that runs prefill and decode, so a store delays the other requests on that engine and a restore delays the next call. As a result, cache capacity alone does not determine whether restoration pays off. In the host-capacity sweep of \S\ref{sec:recovery}, a tier that holds the whole reused set still meets the SLO for only 84.1\% of workflows at the higher arrival rate. This motivates a model that represents engine scheduling explicitly (\S\ref{sec:model}).

\section{System Design}
\label{sec:design}

To answer the two questions above, we build a host tier for the runtime-owned KV of this stack and a model that predicts its serving behavior. As illustrated in Figure~\ref{fig:system-overview}, the design consists of two components: a host-tier KV bridge that keeps evicted planner prefixes in host DRAM and restores them on demand (\S\ref{sec:restore}), and a scheduler-aware discrete-event model whose predictions are frozen before each test cohort (\S\ref{sec:model}). The remainder of this section first gives an overview of the serving path (\S\ref{sec:overview}) and then describes each component.

\subsection{Overview}
\label{sec:overview}

Figure~\ref{fig:system-overview} shows the serving path with a chain workflow as the dependency example; replayed prompts are fixed. Admitted workflows send requests to two engines, a planner that runs with tensor parallelism (TP) over two cards and an executor that runs on a third, and the two engines can overlap in time. On the planner, restore, prefill, and store occupy the same engine thread as decode. Let $H_h$ and $H_l$ denote the usable host and local prefix lengths of a request. When $H_h>H_l$, the full host prefix is restored before the remaining prefill. Missing host entries are stored after prefill, and incomplete stores are invisible to lookups. Both transfers are conditional: a restore runs only when $H_h>H_l$, and a store runs only for missing host entries. Separately from serving, the event model uses the frozen trace and a calibration on seed 701 to make predictions that are fixed before each test cohort. For the store-cost test, only its store coefficient is changed, using the isolated store measurement taken before the serving runs.

\paragraph{Serving policies.} We compare the following serving policies throughout the paper. \emph{B0} is stock serving. \emph{F6} caps active workflows at six and in-flight requests at four per role, admitted in first-in, first-out (FIFO) order. The suffix \emph{+host} adds the host tier of \S\ref{sec:restore} to either policy; from \S\ref{sec:model} on, the host arm is F6+host. Comparing F6 with F6+host holds admission fixed, whereas comparing B0 with F6+host changes both admission and restore.

\subsection{Host-Tier KV Restoration}
\label{sec:restore}

The host tier keeps the KV of evicted planner prefixes in host DRAM and copies it back to the device when a later request can reuse it.

\paragraph{Transfer path.} The runtime exposes KV only through copy calls (\texttt{get\_kv\_cache} and \texttt{set\_kv\_cache}) on its own layout. We mapped the host-visible layout of the serving graph for both TP shards and connected it to vLLM's offloading interface. During serving, prefixes are stored and restored in 128-token slots. A single 2,048-token copy aborted, whereas 512-token copies passed a separate round-trip check and are used only in \S\ref{sec:a0}.

\paragraph{Ordering contracts.} From vLLM's least-recently-used (LRU) offloading manager~\cite{vllmoffload}, we reuse host-block pinning during loads and the store-publication rule that keeps incomplete stores invisible to lookups. Two parts are our own: the mapping from vLLM's slots to the runtime's opaque TP payload, and the link to the runner's synchronous execution boundary. The worker-local bridge reserves device mappings through the scheduler and advances the optimum runner's input past the restored prefix only after both TP payloads have been written. The bridge is synchronous and specific to the single-process executor, and stores run after prefill.

\paragraph{Correctness.} Logits computed from restored KV were bitwise equal to the original on two cards, and restored KV was bitwise equal to recomputed KV for the same graph and input. A 2,048-token round trip took about 165\,ms in a microbenchmark outside serving.

\input{sec_model}

\section{Performance Evaluation}
\label{sec:eval}

We evaluate the host tier and the frozen model on the ATOM+ testbed. After describing the setup (\S\ref{sec:setup}), we examine overload recovery (\S\ref{sec:recovery}), latency and policy ranking at common load (\S\ref{sec:bt}), serving capacity (\S\ref{sec:capacity}), and a change to the store cost (\S\ref{sec:a0}).

\subsection{Evaluation Setup}
\label{sec:setup}

\paragraph{Testbed.} All runs use Rebellions ATOM+ (RBLN-CA22) cards with KMD 3.0.0, compiler and optimum-rbln 0.10.2, and vllm-rbln 0.10.2.post2. The planner, Qwen3-8B~\cite{qwen3} in bf16, runs with tensor parallelism over two cards, and the executor, Qwen3-0.6B, runs on a third.

\paragraph{Workloads.} We replay synthetic workflow DAGs with fixed inputs. A \emph{chain} workflow consists of a planner plan, an executor step after a 0.1\,s tool delay, and a planner answer, with generation budgets of 32 tokens each and 512-token shared prefixes drawn from a pool of 64. A \emph{fork} workflow consists of a planner plan with a 32-token budget, two parallel executor branches with 128-token budgets after a 1\,s tool delay, and a planner answer with a 64-token budget, with 2,048-token prefixes drawn from a pool of 128. Budgets are maximum generation lengths, and requests stop at the natural end of sequence. We also use \emph{mixed} (50/50 chain and fork) and \emph{mixed-burst} (mixed with arrivals at twice the mean rate for 15\,s and two-thirds of it for 45\,s in every 60\,s cycle). Arrivals are open loop. Generated outputs do not change later prompts, so the replay measures serving behavior and does not assess agent quality. A 50,000-workflow dataset in the same four profiles, together with the trace generator, is publicly available~\cite{woo2026replay}.

\paragraph{Metrics.} The SLOs are 30\,s per chain workflow and 120\,s per fork workflow. SLO attainment uses all offered workflows, latency quantiles use completed workflows, and throughput counts successful completions over elapsed time including drain. Errors are always reported.

\paragraph{Cohorts.} Table~\ref{tab:cohorts} lists the runs used. Each cohort keeps its own frozen protocol, and we do not pool them. For the three test cohorts, predictions were written to a file and hashed before the first run. Error definitions follow each cohort's frozen protocol and differ: the common-load cohort uses $(\hat y-y)/y$, whereas the endpoint and store cohorts use $|y-\hat y|/\hat y$. We keep them separate instead of merging them into one hit rate.

\begin{table}[t]
\centering\small
\setlength{\belowcaptionskip}{6pt}
\caption{Cohorts. Runs have 1,200 workflows unless noted. The collapse cohort also includes 48-workflow observer checks, screen probes have at least 480 workflows and 600\,s of arrivals, and confirmation runs have at least 1,200 workflows and 1,800\,s (more at high rates). Only the calibration cohort is used to fit the model. Client: original, or with the event-loop fix described in \S\ref{sec:setup}.}
\label{tab:cohorts}
\begin{tabular}{@{}llll@{}}
\toprule
Cohort & Seeds & KV & Client \\
\midrule
collapse lineage & 501 & 64 & orig. \\
restore, fork 0.55 & 501 & 64 & orig. \\
restore, burst 0.65 & 505 & 32 & orig. \\
host capacity $\times$ load & 506 & 32 & orig. \\
calibration (480 wf) & 701 & 32 & orig. \\
common load, 24 cond. & 702, 703 & 32 & fixed \\
capacity screen & 711 & 32 & fixed \\
endpoint confirmation & 712, 713 & 32 & fixed \\
store-cost change & 704, 705 & 32 & fixed \\
\bottomrule
\end{tabular}
\end{table}

\begin{figure*}[t]
\centering
\includegraphics[width=\textwidth]{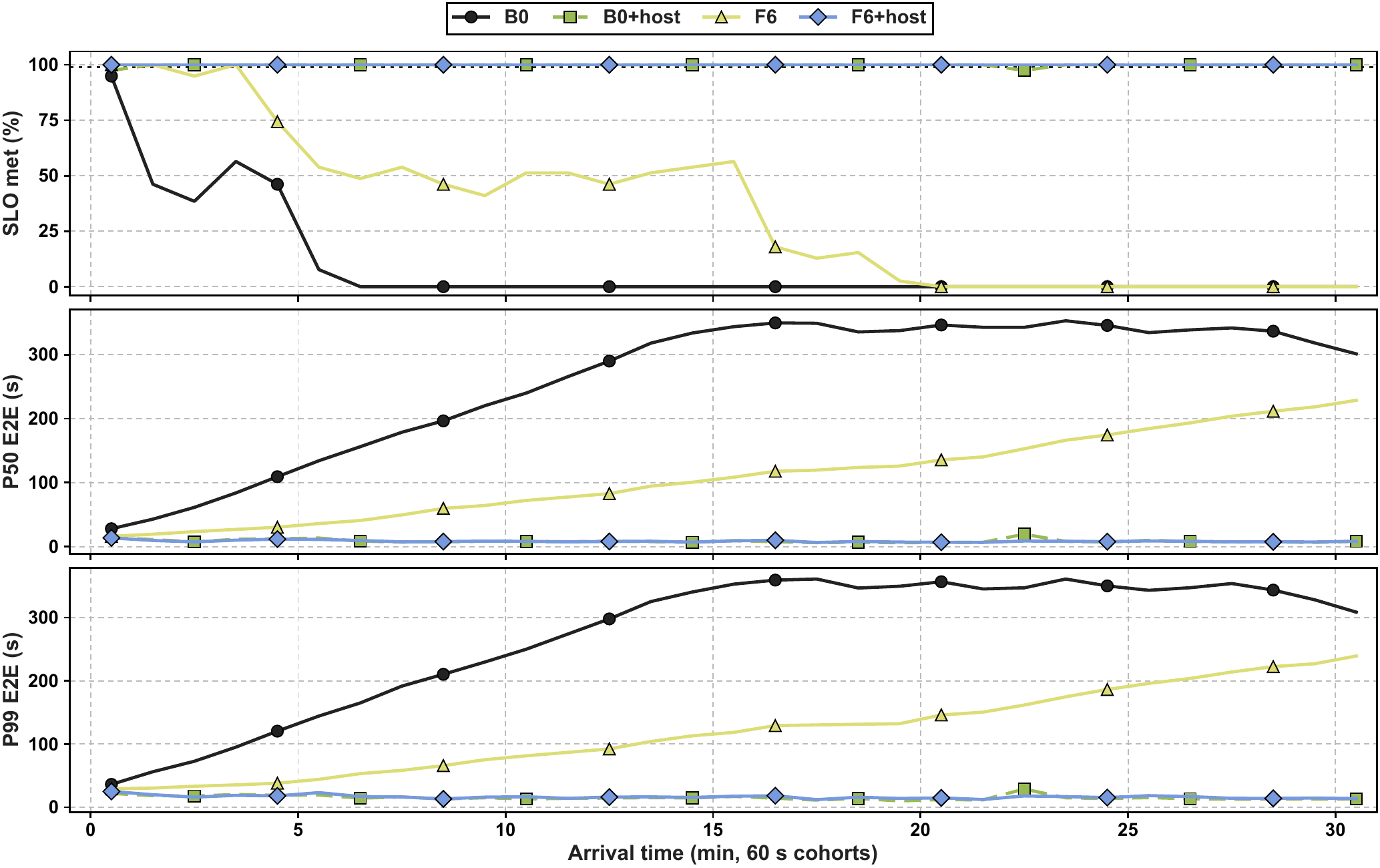}
\caption{Recovery under KV32 mixed-burst overload (seed 505, 0.65 workflows/s, 1,200 arrivals per arm). Each point summarizes a 60\,s arrival cohort; the horizontal axis is arrival time, not completion time. SLO attainment counts all workflows including errors, and latency quantiles use successful workflows only. With the host tier, both B0 and F6 keep SLO attainment near 100\% and latency flat, whereas without it SLO attainment falls to zero and latency rises to hundreds of seconds. Error counts are in Table~\ref{tab:recovery}.}
\label{fig:timeline}
\end{figure*}

\paragraph{Measurement client.} Five connection closures in early runs were consistent with an idle keep-alive reuse race (the server closed idle connections after 5\,s; the client could reuse them for up to 15\,s). We reproduced the race in a CPU-only test but did not trace the exact cause of the five closures. We switched all arms to one fresh connection per request, without retries. A crossover of pooled and fresh connections did not establish a consistent transport effect on tail latency (block P99 ratios 0.900 and 2.416), but it exposed a 5.2\,s client event-loop stall. Synchronous file I/O was a candidate, so later cohorts moved file operations off the loop and passed a separate I/O check. The first five cohorts in Table~\ref{tab:cohorts}, including calibration, used the original client, and the four test cohorts used the fixed one. The model was therefore calibrated before the fix, and we did not model the fix's effect. Earlier cohorts are reported as measured.

\input{sec_energy}

\subsection{Overload Recovery}
\label{sec:recovery}

\begin{figure*}[t]
\centering
\includegraphics[width=\textwidth]{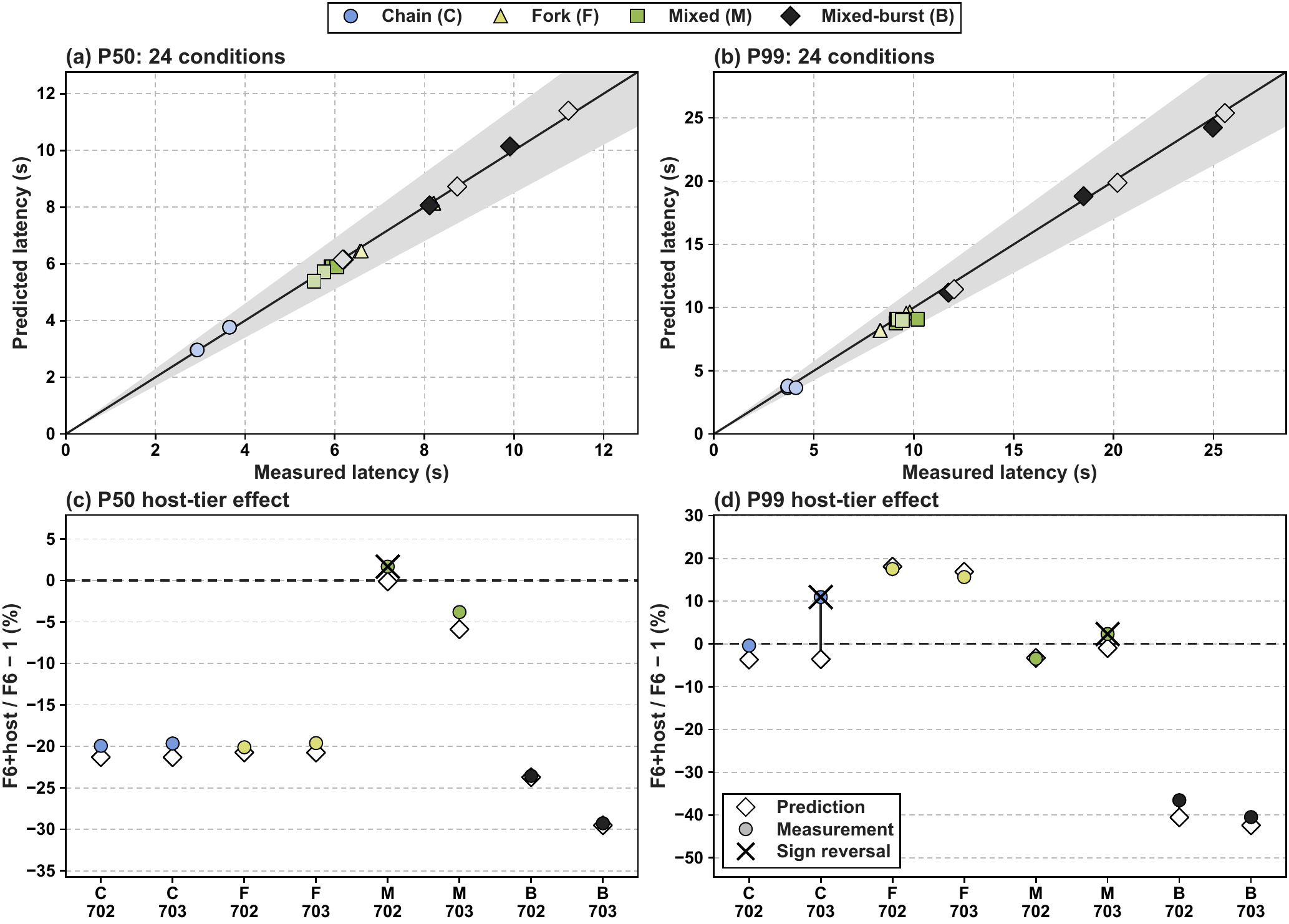}
\caption{Frozen common-load forecasts against measurements for 24 conditions (top) and predicted against observed host-tier effects at fixed F6 admission (bottom). Every P50 and P99 forecast falls within the frozen 15\% tolerance (shaded band, not an uncertainty interval), while crosses mark the three host-tier effects whose sign was mispredicted. In the top panels, dark and pale fills are seeds 702 and 703.}
\label{fig:bt}
\end{figure*}

We first test whether the host tier recovers the overload conditions in which serving without it collapses. Table~\ref{tab:recovery} and Figure~\ref{fig:timeline} show the two overload conditions. With the host tier, the server absorbed nearly the whole arrival rate in all four cases, and summed planner prefill API time fell from 1,115--1,708\,s to 262--265\,s across the four comparisons (host-inclusive, warmup included).

\begin{table}[t]
\centering\small
\setlength{\belowcaptionskip}{6pt}
\caption{Workflow SLO attainment without $\to$ with the host tier, 1,200 workflows, one run each. Errors are failed workflows.}
\label{tab:recovery}
\setlength{\tabcolsep}{3pt}
\begin{tabular}{@{}lllll@{}}
\toprule
Condition & Policy & SLO (\%) & Errors & wf/s \\
\midrule
KV64 fork 0.55 & B0 & 22.8$\to$100 & 536$\to$0 & 0.269$\to$0.549 \\
 & F6 & 73.8$\to$100 & 0$\to$0 & 0.516$\to$0.549 \\
KV32 burst 0.65 & B0 & 9.4$\to$99.8 & 291$\to$2 & 0.429$\to$0.650 \\
 & F6 & 34.8$\to$100 & 0$\to$0 & 0.581$\to$0.651 \\
\bottomrule
\end{tabular}
\end{table}

However, the recovery does not hold at every load. In a host-capacity sweep (F6, KV32 mixed-burst, seed 506), 9\,GiB was enough at 0.65 workflows/s (SLO attainment of 99.9\% against 28.7\% without the tier), but at 0.80 workflows/s SLO attainment was 21.1\%, 40.4\%, and 84.1\% with 9, 18, and 36\,GiB, although the reused set fit in 36\,GiB. Capacity alone therefore did not decide the outcome, which is why the model of \S\ref{sec:model} represents engine scheduling.

\subsection{Latency and Policy Ranking at Common Load}
\label{sec:bt}

\begin{figure*}[t]
\centering
\includegraphics[width=\textwidth]{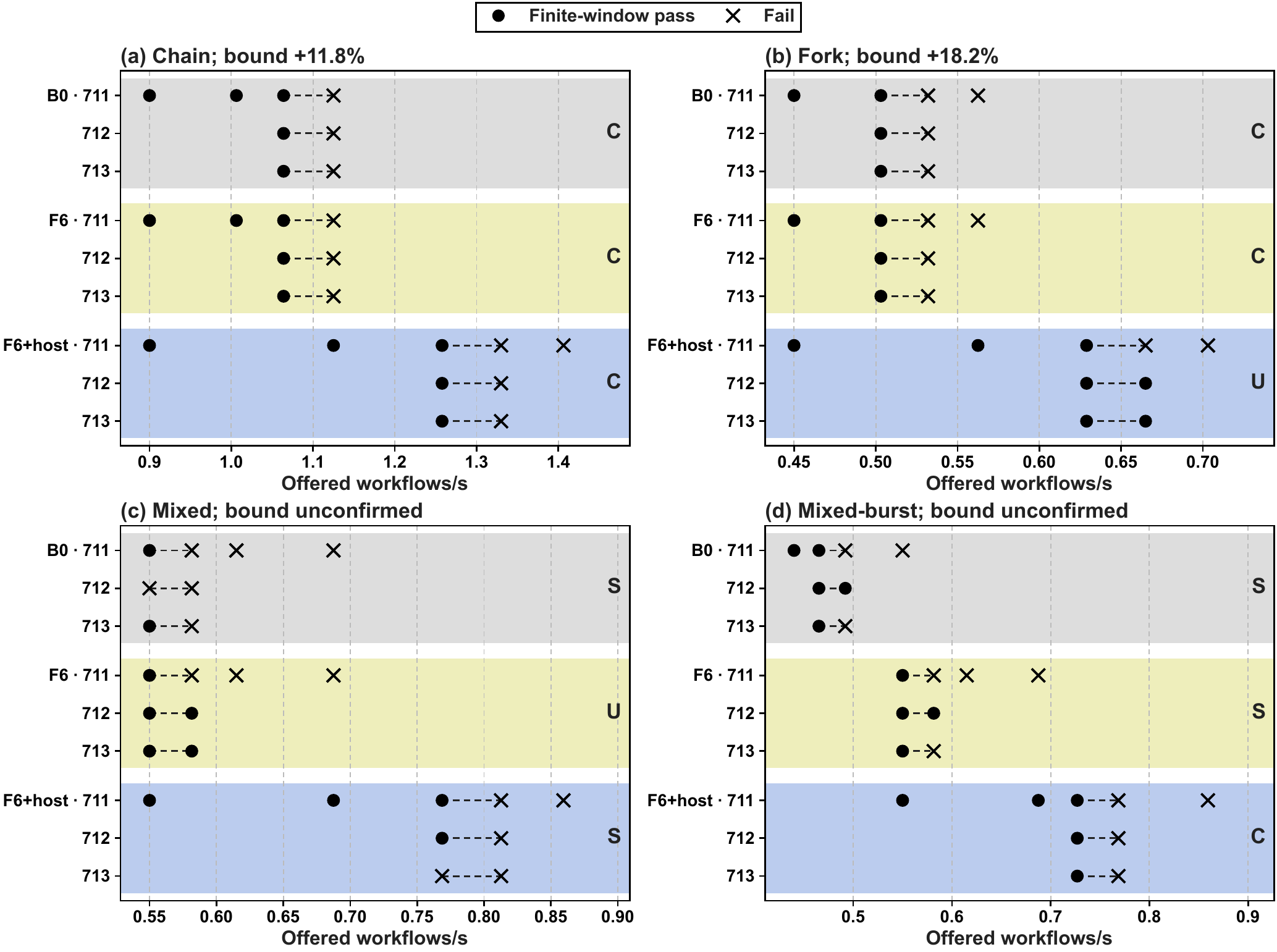}
\caption{Capacity screen (seed 711, all 52 probes) and frozen endpoint confirmation (seeds 712/713, 48 runs) per workload and policy. Filled circles are finite-window passes and crosses are failures; dashed spans are brackets, not confidence intervals. C: the low endpoint passed and the high endpoint failed on both new seeds; U: both endpoints passed; S: seed-sensitive. The host-tier gain is confirmed for chain and fork only. Bounds in the panel titles are the frozen host-tier low endpoint over the B0 high endpoint, minus one.}
\label{fig:capacity}
\end{figure*}

We next examine a load that every policy sustains, which separates latency prediction from capacity. The common-load cohort covers \{B0, F6, F6+host 36\,GiB\} $\times$ \{chain, fork, mixed, mixed-burst\} $\times$ two seeds, with 1,200 workflows each, at 0.8 times the calibration rates (chain 0.72, fork 0.36, mixed 0.44, and burst 0.44 workflows/s). Predictions were frozen before the first run with a 15\% tolerance.

\paragraph{Prediction accuracy.} All 24 runs had zero errors and 100\% SLO attainment. Because B0 completes every condition, this cohort tests latency prediction rather than capacity. Every predicted P50 and P99 was within tolerance, and the largest errors were 3.1\% (P50) and 11.0\% (P99).

\begin{table}[t]
\centering\small
\setlength{\belowcaptionskip}{6pt}
\caption{F6 $\to$ F6+host at common load, seed 702 / 703. Last column: B0 $\to$ F6+host, so it includes the admission cap as well as the tier; gross board energy over the fixed offered window per offered workflow.}
\label{tab:bt}
\setlength{\tabcolsep}{3pt}
\begin{tabular}{@{}lccc@{}}
\toprule
Workload & P50 (\%) & P99 (\%) & J/wf vs.\ B0 (\%) \\
\midrule
chain & $-19.9$ / $-19.7$ & $-0.4$ / $+11.0$ & $-3.7$ / $-3.9$ \\
fork & $-20.1$ / $-19.6$ & $+17.5$ / $+15.6$ & $-6.5$ / $-6.9$ \\
mixed & $+1.7$ / $-3.8$ & $-3.4$ / $+2.3$ & $-6.1$ / $-6.3$ \\
burst & $-23.6$ / $-29.3$ & $-36.5$ / $-40.5$ & $-10.5$ / $-10.4$ \\
\bottomrule
\end{tabular}
\end{table}

\paragraph{Effect of the host tier.} Table~\ref{tab:bt} shows the host effect against F6, and two effects are large. Under bursts, restore cut P99 by 37--40\%. For fork workflows, in contrast, it raised P99 by 16--18\% while cutting P50 by 20\%. In the model, this increase comes from synchronous stores on the planner engine: with the store cost set to zero, it disappears (\S\ref{sec:a0}). The model predicted both effects in sign and size, and the fork loss was in the frozen file before the runs.

\paragraph{Policy ranking.} Three policy contrasts had the wrong sign. Against F6, the model predicted $-3.58\%$ for chain P99 on seed 703 and the run showed $+10.96\%$; for mixed, it predicted $-0.12\%$ (P50, seed 702) and $-0.97\%$ (P99, seed 703) against $+1.67\%$ and $+2.32\%$. The absolute errors were within tolerance in all three, but the ranking was wrong, and in the chain case the observed difference was sizable (Figure~\ref{fig:bt}). At common load, the model thus predicted absolute latency well but did not reliably rank the policies.

\paragraph{Output differences.} Throughput agreement carries little information about capacity, because open-loop throughput equals the arrival rate whenever the system keeps up. The generated outputs also differ between arms. Comparing token sequences request by request, the host arm differed from B0 in 197 and 172 of 3,600 chain requests and in 450 and 414 of 4,800 fork requests. Restore was bitwise exact in isolated checks. In these runs, however, the arms also differ in admission and timing, and so in which requests share a decode batch and which compiled batch shape runs; we have not attributed the differences to any of these factors.

\subsection{Serving Capacity}
\label{sec:capacity}

We then ask how much additional load the host tier supports. Because a finite run cannot prove stability, we define a pass/fail judge, screen each policy and workload for a load bracket on one seed, and confirm both endpoints of every bracket on two new seeds.

\paragraph{Finite-window judge.} A run passes if it has zero errors, at least 99\% SLO attainment in each workflow class, bounded backlog, and a short drain. Backlog $Q(t)$ is released minus finished workflows, sampled over the 20--90\% part of the arrival window (at cycle boundaries for bursts). Both the growth between the first and last sevenths of the samples and the regression slope times the window must stay below $\max(6, 0.01N)$ for $N$ offered workflows, and the drain after the last arrival must stay below $\max(20\,\text{s}, 0.03T)$. A finite window cannot prove stability, but the screen does reject runs that meet the SLO only because the window ended: one chain run at 1.125 workflows/s had 100\% SLO attainment and a backlog growth of 13 workflows.

\paragraph{Bracket screen.} With seed 711 and 480 or more workflows per probe, each policy and workload started at the calibration rate, moved by $\times1.25$ after a pass and $\times0.8$ after a failure, and then bisected geometrically until the bracket ratio was at most 1.10 or eight probes were used. All 12 brackets closed monotonically in 52 probes.

\paragraph{Endpoint confirmation.} Both endpoints of all 12 brackets were rerun on seeds 712 and 713 with at least 1,200 workflows and 30 minutes of arrivals (48 runs), with the model's pass/fail calls frozen beforehand. We claim a gain only where the host arm's lower endpoint passed on both seeds and the baseline's upper endpoint failed on both, and then report $L_\text{host}/U_\text{base}-1$.

\begin{table}[t]
\centering\small
\setlength{\belowcaptionskip}{6pt}
\caption{Confirmed capacity endpoints (workflows/s). $\checkmark$: pass on both new seeds; $\times$: fail on both; mixed: one of each.}
\label{tab:capacity}
\begin{tabular}{@{}lccc@{}}
\toprule
Workload & Host lower & B0 upper & Gain \\
\midrule
chain & 1.258 $\checkmark$ & 1.125 $\times$ & $\ge$11.8\% \\
fork & 0.629 $\checkmark$ & 0.532 $\times$ & $\ge$18.2\% \\
mixed & 0.769 mixed & 0.582 $\times$ & not established \\
burst & 0.727 $\checkmark$ & 0.492 mixed & not established \\
\bottomrule
\end{tabular}
\end{table}

\begin{figure*}[t]
\centering
\includegraphics[width=\textwidth]{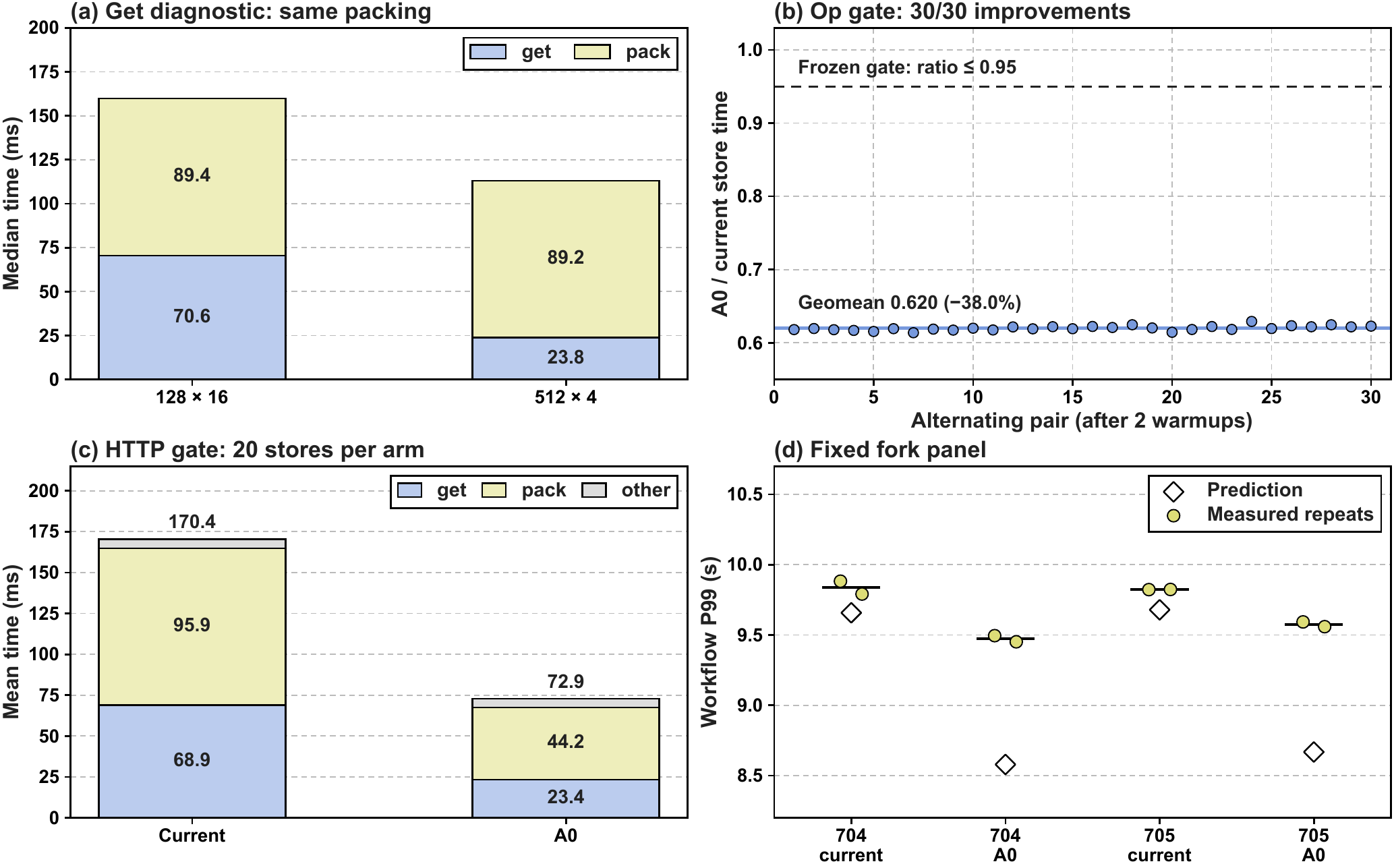}
\caption{Store path before and after the change (A0). (a) Copy-size diagnostic with identical packing; (b) the op gate, isolated total store time excluding pool setup (7.8\,s); (c) the HTTP gate, serialized requests with copy and packing timed separately; (d) the serving panel, measured repeats against frozen predictions. A0 cuts isolated store time by 38\%, but the measured fork P99 improves far less than predicted. The panels are separate measurements and are not combined.}
\label{fig:store}
\end{figure*}

\paragraph{Results.} Table~\ref{tab:capacity} gives the result. In chain and fork, F6's upper endpoint also failed on both seeds, so the bound against F6 is the same. In mixed, the host arm's lower endpoint missed on seed 713 with 98.8\% SLO attainment. Under bursts, B0's upper endpoint passed on one seed. The screen's single-seed brackets suggested gains of 32\% and 48\% for these two workloads, but the confirmation does not support them. The fork host arm passed its \emph{upper} endpoint (0.665) on both new seeds, so 18.2\% is conservative, but recomputing the bound from that point would be a post hoc choice.

Near the boundary, outcomes are sensitive to the seed. B0 in mixed collapsed at its lower endpoint on seed 712 (SLO attainment of 27.6\%) and passed on seed 713. The frozen model's calls matched 45 of the 48 outcomes. Predicting a pass at every lower endpoint and a failure at every upper endpoint, as the screen labeled them, would have matched 40. On the nine runs where the model and the screen disagreed, the model was right seven times (Figure~\ref{fig:capacity}).

\subsection{Store-Path Optimization}
\label{sec:a0}

Finally, we test the explanation of the fork tail loss. If the loss came from synchronous stores, making stores cheaper should remove it. The model supports this hypothesis: setting the store cost to zero brought fork P99 back to the level without the tier, while setting the restore cost to zero did not change it. In the measured runs, the host arm's P99 was 9.9\,s in the first half and 6.8--7.3\,s in the second, consistent with a burst of initial stores.

\paragraph{Pooled store path.} Timers added during the capacity screen split store time into a copy from the device (40--42\%) and host-side packing (55--57\%). We then built an alternative store path, A0 (the existing path is \emph{current}): a 36\,GiB pool of fixed slots allocated and touched before measurement and reused after eviction, copies of 512 instead of 128 tokens, and packing straight into pool slots. When the pool is exhausted, the stock skip-store behavior is kept and counted. Pool setup took 7.8\,s at server start.

\paragraph{Isolated gates.} Before the serving runs, we tested A0 in two isolated settings, which Figure~\ref{fig:store} labels the op gate and the HTTP gate. On a fixed 2,048-token source, all correctness controls passed (byte equality, continuation, poison and load-pin checks, and 21 serialized HTTP requests with restore and identical outputs). In the op gate, the foreground store time fell from 119.7 to 74.1\,ms (ratio 0.620, 30 of 30 pairs). In the HTTP gate, where the steps were timed separately, the copy fell from 68.9 to 23.4\,ms and packing from 95.9 to 44.2\,ms per 2,048 tokens (store 170.4 to 72.9\,ms). The copy gain matches a per-call overhead: the same 2,048 tokens took 70.6\,ms in 128-token calls and 23.8\,ms in 512-token calls. Preallocation, buffer reuse, and the copy routine changed together, so we cannot separate their shares of the packing gain. A CPU microbenchmark on another server showed a 3.5$\times$ packing gain from preallocated output buffers alone, but we do not carry that number over.

\paragraph{Serving panel.} The serving panel is fork at 0.36 workflows/s on seeds 704 and 705, with four runs per arm in crossover order. Predictions were frozen after the op gate by scaling only the store coefficient by the gate ratio. Every run passed, and every metric was within tolerance. However, the arithmetic mean of the four run-level P99 values per arm fell from 9.83 to 9.53\,s ($-3.1\%$; $-3.7\%$ and $-2.5\%$ by seed), while the model predicted 9.66--9.68 to 8.58--8.67\,s ($-11.2\%$ and $-10.4\%$). The model was close for the current path (errors 1.4--2.3\%) and consistently optimistic for the new one (10.2--10.7\%; Figure~\ref{fig:store}d).

\paragraph{Tail requests.} We followed the 12 slowest workflows of every panel run through the lineage and attributed each planner request's gaps between decode steps to whatever the engine was running. On this planner-only denominator, other requests' prefill covered 75.8--86.5\% of the gap time in the panel runs. In the common-load runs, it covered 89.6\% and 89.4\% without the tier and 79.1\% and 80.2\% with it. None of the slowest workflows restored its own prefix, and restores never overlapped the prefill timers. The slowest workflows differ between arms, however: over the union of slowest IDs per seed, prefill exposure was 28.6$\to$26.4\,s and 28.5$\to$27.4\,s from the arm without the tier to the arm with it, with store and restore exposure appearing only in the latter. These are request-seconds, so overlapping waiters count more than once, and they are not removable critical-path time. In both arms, tail planner requests spent most of their waiting time behind other prefill calls. That shared cost cannot explain the difference between the arms, and the store and restore exposure is not critical-path time, so the host-minus-baseline P99 difference remains unattributed. The frozen model overpredicted the P99 benefit of the cheaper store.

\paragraph{Interleaving prefill.} Long prefills already run as 128-token chunks, so we asked whether decode steps could be placed between them. The installed scheduler admits one prefill at a time and gives it the engine, and the runner rejects mixed prefill and decode batches. The runtime itself allowed alternating 128-token prefill calls of one request with decode calls of another, with bitwise-identical logits in three repetitions. In the model, interleaving at 128, 256, or 512 tokens with zero switching cost gave fork P99 of 10.34, 9.93, and 9.80\,s against 9.66\,s, because faster decode for running requests delays new requests' prefill. Because the same model overpredicted the A0 benefit, this result is only weak evidence about the hardware, and we did not pursue a scheduler change.

\section{Discussion}
\label{sec:discussion}

\paragraph{Scope of the measurements.} We measured one vendor stack, one SDK version, and one fixed heterogeneous placement. The DAGs are synthetic with fixed inputs: outputs do not change later prompts, so agent semantics, tool choice, and execution paths are not validated. The planner's unique reuse set (36\,GiB) fits the largest host tier we used; the 9/18/36\,GiB sweep shows what happens when it does not. Host DRAM use exceeds the payload: the anonymous-memory peak was about 85\,GiB in one cohort, and the pool of \S\ref{sec:a0} is resident from server start.

\paragraph{Measurement limitations.} Energy is gross board power over a fixed offered window (\S\ref{sec:setup}), and host CPU and DRAM energy were not measured. Output differences are counted per request and not attributed to a cause. The measurement client changed between cohorts as described in \S\ref{sec:setup}. The simulator executes the installed SDK's block-management code, so it is specific to this SDK version.

\section{Related Work}

\paragraph{Host-tier KV.} LMCache~\cite{lmcache2025} keeps KV outside the engine and moves it between tiers. SGLang HiCache~\cite{hicache2025} separates host and device layouts, overlaps transfers, and chooses write policies. vLLM's offloading connector and LRU manager~\cite{vllmoffload} define the not-ready, pin, and publish rules we reuse. InferCept~\cite{abhyankar2024infercept} studies whether to preserve, discard, or swap context during tool calls and pipelines its swaps.

\paragraph{Vendor offloading.} The 0.12 documentation of Rebellions describes an offloading path based on LMCache. The separate \texttt{lmcache-rbln} package supplies the device-specific integration, and public vllm-rbln code registers and re-exports its connector. The documented path requires the \texttt{vllm} model implementation and KV tensors on the RBLN device, and supports host-memory and direct-NVMe tiers~\cite{rblnoffload}. Our measurements use the pinned 0.10.2 optimum path, whose runtime owns the KV allocation. We bridge its public get/set interface to vLLM's offloading machinery so that this older runtime can be measured with a synchronous host-DRAM tier. We did not evaluate the vendor connector, and whether our numbers transfer to it is untested.

\paragraph{Agent serving.} Several systems schedule or cache by program structure, including Parrot~\cite{lin2024parrot}, Autellix~\cite{luo2025autellix}, and KVFlow~\cite{pan2025kvflow}. CONCUR~\cite{chen2026concur} and ThunderAgent~\cite{kang2026thunderagent} control admission to avoid thrashing, and our F6 policy is a fixed cap of that kind. In contrast to these systems, our study adds measurements on a static-graph stack: what work a collapse consists of, what restoring it costs in the foreground, and how well a model fixed in advance predicted the result.

\section{Conclusion}

In the two tested overload conditions, a host tier replaced repeated planner prefill with KV restoration and recovered most or all offered workflows. Separate endpoint confirmations supported conservative load gains for chain and fork, while mixed and burst endpoints remained seed-sensitive. At common load, the host tier improved median latency but worsened the fork tail. A frozen model anticipated several of these effects, yet it misordered three policy contrasts and overpredicted the benefit of cheaper stores. The measured trade-off depends on runtime scheduling and transfer cost as well as cache capacity.

\section*{Acknowledgments}
This research was supported by the Ministry of Science and ICT (MSIT) through the 2026 High-Performance Computing (HPC) Support Program (RQT-26-041563).

\bibliographystyle{plain}
\bibliography{refs}

\appendix
\input{app_model}

\end{document}

%% file: figs/fig_overview.tex
\definecolor{ovGray}{HTML}{DDDDDD}
\definecolor{ovSand}{HTML}{EEEEBB}
\definecolor{ovBlue}{HTML}{BBCCEE}
\definecolor{ovCyan}{HTML}{CCEEFF}
\definecolor{ovSage}{HTML}{CCDDAA}
\definecolor{ovPanel}{HTML}{F7F7F7}
\definecolor{ovText}{HTML}{222222}
\begin{tikzpicture}[
  x=0.985cm, font=\fontfamily{phv}\fontseries{b}\selectfont\boldmath\scriptsize,
  text=ovText,
  region/.style={draw=black, line width=0.6pt, rounded corners=3pt},
  box/.style={draw=black, line width=0.45pt, rounded corners=2pt, fill=white, align=center, inner sep=2pt, text width=#1, minimum height=0.6cm},
  box/.default=2.4cm,
  title/.style={font=\fontfamily{phv}\fontseries{b}\selectfont\boldmath\footnotesize},
  flow/.style={-{Stealth[length=4pt,width=3.6pt]}, line width=0.8pt},
  thin/.style={-{Stealth[length=3.5pt,width=3.2pt]}, line width=0.6pt},
  lab/.style={inner sep=1pt},
]
\draw[region, fill=ovGray] (0,5.25) rectangle (6.0,6.6);
\node[title, anchor=north] at (3.0,6.55) {Synthetic workflow DAG (chain)};
\node[box=1.25cm, fill=ovCyan] (w1) at (0.95,5.68) {Planner\\plan};
\node[box=1.25cm, fill=ovSage] (w2) at (3.0,5.68) {Executor\\step};
\node[box=1.25cm, fill=ovCyan] (w3) at (5.05,5.68) {Planner\\answer};
\draw[thin] (w1) -- (w2);
\draw[thin] (w2) -- (w3);

\draw[region, fill=ovSand] (6.5,5.25) rectangle (17.55,6.6);
\node[title, anchor=north] at (12.025,6.55) {F6 admission};
\node[align=center] at (12.025,5.68) {at most six active workflows \quad$\cdot$\quad at most four in-flight requests per role \quad$\cdot$\quad FIFO order};
\draw[flow] (6.0,5.92) -- (6.5,5.92);

\draw[region, fill=ovCyan] (0,1.55) rectangle (11.55,4.85);
\node[title, anchor=north west, align=left] at (0.15,4.8) {Planner\\[-1pt]{\scriptsize Qwen3-8B, two cards (TP2)}};
\node[box=2.6cm] (pq) at (7.4,4.28) {FIFO request queue};
\draw[flow] (7.4,5.25) -- (pq.north);

\draw[line width=0.45pt, rounded corners=2pt, fill=white] (0.25,2.6) rectangle (11.3,3.82);
\node[anchor=north west, inner sep=1.5pt] at (0.3,3.8) {Same engine thread};
\node[box=2.45cm, fill=ovBlue] (r) at (1.65,3.0) {Restore host prefix\\if $H_h{>}H_l$};
\node[box=2.45cm, fill=ovPanel] (pf) at (4.45,3.0) {Remaining prefill\\128-token chunks};
\node[box=2.45cm, fill=ovBlue] (st) at (7.25,3.0) {Store missing\\host entries};
\node[box=2.0cm, fill=ovPanel] (dc) at (9.85,3.0) {Decode};
\draw[thin] (r) -- (pf); \draw[thin] (pf) -- (st); \draw[thin] (st) -- (dc);
\draw[flow] (pq.south) -- (7.4,3.82);

\node[box=3.3cm] (kv) at (2.0,1.98) {Runtime-owned device KV\\(both TP payloads)};
\node[box=3.6cm, fill=ovBlue] (host) at (9.45,1.98) {Host DRAM tier\\(incomplete stores invisible)};
\draw[thin] ([yshift=4pt]kv.east) -- node[above, lab] {store (\texttt{get\_kv\_cache})} ([yshift=4pt]host.west);
\draw[thin] ([yshift=-4pt]host.west) -- node[below, lab] {restore (\texttt{set\_kv\_cache})} ([yshift=-4pt]kv.east);

\draw[region, fill=ovSage] (12.0,1.55) rectangle (17.55,4.85);
\node[title, anchor=north west, align=left] at (12.15,4.8) {Executor\\[-1pt]{\scriptsize Qwen3-0.6B, one card}};
\node[box=2.6cm] (eq) at (15.6,3.71) {FIFO request queue};
\draw[flow] (15.6,5.25) -- (eq.north);
\draw[line width=0.45pt, rounded corners=2pt, fill=white] (12.25,2.44) rectangle (17.3,3.23);
\node[box=1.75cm, fill=ovPanel] (epf) at (13.65,2.835) {Prefill};
\node[box=1.75cm, fill=ovPanel] (edc) at (15.9,2.835) {Decode};
\draw[thin] (epf) -- (edc);
\draw[flow] (eq.south) -- (15.6,3.23);
\node[box=3.3cm] (ekv) at (14.775,1.95) {Runtime-owned device KV};

\draw[region, dashed, fill=ovPanel] (0,0) rectangle (17.55,1.2);
\node[title, anchor=north west] at (0.15,1.17) {Prospective evaluation (does not control serving)};
\node[box=3.4cm] (m1) at (2.0,0.42) {Frozen DAG trace +\\seed-701 calibration};
\node[box=3.4cm, fill=ovGray] (m2) at (6.6,0.42) {Reduced discrete-event\\model};
\node[box=3.4cm, fill=ovSand] (m3) at (11.2,0.42) {Predictions frozen\\before each cohort};
\node[box=3.6cm] (m4) at (15.55,0.42) {Common load, capacity\\endpoints, store-cost change};
\draw[thin] (m1) -- (m2); \draw[thin] (m2) -- (m3); \draw[thin] (m3) -- (m4);
\end{tikzpicture}

%% file: sec_model.tex
\subsection{Scheduler-Aware Model}
\label{sec:model}

The second component predicts the serving behavior of a policy before it is run. We first explain why the model must simulate the engine queues and then describe its rules, calibration, and use.

\paragraph{Why occupancy is an output.} Our first capacity forecast replayed the cache to predict host hit rates and combined them with mean call costs. In the host-capacity sweep of \S\ref{sec:recovery}, hit rates were right within 1.5 percentage points at all eight points. Without the tier, the predicted completion rate was 0.5808 workflows/s against 0.5880 measured (finite-run throughput including drain). With the tier, however, the completion rate was under-predicted by 16--17\% with 9 and 18\,GiB at 0.80 workflows/s, beyond the 15\% tolerance. Per-call decode time stayed at 35.6--36.9\,ms while the average number of tokens per decode call rose from 2.6 to 3.3--3.7 under load, which accounts for part of the residual. A model that fixes occupancy cannot follow this change.

\paragraph{Event model.} We model the vendor scheduler with a reduced discrete-event simulation. Each role has a FIFO request queue, at most four running requests, and one busy flag, and the two role engines can overlap. B0 admits workflows without a client cap, whereas F6 admits six workflows and four requests per role. From here on, the candidate is F6 with the 36\,GiB host tier (F6+host); B0 with the tier alone is not carried forward. Dependencies and tool delays release subsequent nodes, and equal-time events execute in insertion order.

An idle engine first tries to allocate the oldest waiting request. If allocation succeeds and a running slot is free, it serves that request's entire remaining prefill before any decode. Otherwise, it decodes one token for every running request. Prefill costs the required number of 128-token calls plus one wrapper and dispatch overhead, and decode cost is looked up by role and active request count. Batch occupancy therefore follows the simulated queues rather than an assumed function of load.

\paragraph{Cache model.} The cache model reuses the installed outer-block mapping, search, and eviction code, with a reduced inner-block model. A planner-only host directory tracks prefix residency in LRU order. A restore uses the full host prefix when it exceeds local coverage. Restore, prefill, and store occupy the same engine, and new host entries become visible only at completion. These are timing and metadata operations; no tensors or model outputs are simulated.

\paragraph{Calibration.} We pool eight calibration runs: four workload families, each with B0 and F6+36\,GiB, all at seed 701. Each has 480 measured and 32 warmup workflows. Successful call spans provide mean runtime costs, wrapper residuals, and dispatch gaps. Response-token counts provide role/node/context length distributions, capped by the request budget. In the calibration run, every response reached its budget, so each of these distributions is a single value. No held-out timings or output lengths enter calibration, and costs include host-side execution. We fit no end-to-end latency correction and do not prescribe decode occupancy. In the in-sample check, the final calibration still underpredicts fork-B0 P99 by 16.0\%.

\paragraph{Frozen predictions.} The model and coefficients remain fixed for the common-load and endpoint tests. For the store-cost test, only the store coefficient is scaled, by the ratio measured in the isolated store test, before the serving runs. Endpoint predictions use the same finite-window judge as measurements, but simulate neither request errors and timeouts nor preemption or rare stalls. Appendix~\ref{app:model} gives the event rules, coefficients, and cohort-specific freezes.

%% file: sec_energy.tex
\paragraph{Energy.}
We sample \texttt{rbln-stat --json} approximately once per second and sum gross board power over the three occupied physical cards. We identify these cards from runtime contexts after model loading, because logical device indices need not match physical IDs; the common-load and store-cost runs used physical cards 0, 1 and 3. Sample times are the midpoint of each query on the monotonic clock. We integrate linearly interpolated power by the trapezoidal rule, without idle subtraction. The primary comparison window starts at the first measured arrival and ends at the last scheduled arrival plus the largest declared workflow SLO. Models and logging remain active through that endpoint even if work finishes early. Energy per offered workflow divides this integral by the fixed offered count. We do not extrapolate missing boundary coverage. Equal-throughput comparisons additionally require complete work, zero errors, SLO attainment and the declared stability check. Host CPU and DRAM power, the unused card and facility overhead are excluded, so the savings are board-level only.

%% file: app_model.tex
\section{Model Specification and Calibration}
\label{app:model}

\subsection{Inputs, Queues, and Completion}

The input is the frozen tokenized DAG trace: arrival offsets, role assignments, dependencies, tool delays, generation limits and workflow SLOs. Subsequent prompts are fixed by this trace; generated text does not change them. The event heap is ordered by time and then a monotonically increasing insertion counter. Each simulation uses random seed 1701. It first drains 32 warmup workflows, retains both cache states, and starts measured arrivals 7.2\,s later, which approximates the client's 6\,s wait and 1.2\,s initial telemetry interval.

Workflows enter a FIFO admission queue. B0 has no client-side workflow or role cap. F6 permits six active workflows and four in-flight requests per role; excess requests wait FIFO. Both policies use engines with at most four running requests. Ready nodes are visited in trace order. Their output lengths are drawn when dependencies become ready, then a send event is scheduled after the declared tool delay. A node completion releases its role slot before making dependent nodes ready. A workflow completion releases its admission slot. Engine completions defer these callbacks to same-time events, so they cannot interrupt a partially processed completion.

The planner and executor have independent busy flags. A busy engine does nothing when pumped. An idle engine attempts prefill only for its oldest waiting request, and only if fewer than four requests are running and cache allocation succeeds. It does not bypass an unallocatable head request. If that attempt fails, it decodes the current running set. Prefill produces the first output token; each decode completion adds one token to each member of the captured batch. Finished requests release their native allocation. A decode allocation that would require preemption is reported as an error, as are deadlock and exceeding a 100,000\,s simulation limit.

\subsection{Cache State and Publication}

Each role has 256 inner blocks of 128 tokens and eight outer blocks of 4,096 tokens, with maximum model length 8,192 and four sequences. We execute the preserved 0.10.2 outer-block mapping, search and FIFO eviction implementation. Unused tensor allocation and logging imports are stubbed, and the rest of the scheduler is not executed. The reduced inner model maintains a FIFO free-ID queue and a directory that can contain duplicate hashes. Allocation removes old hash entries for reused IDs and installs complete prompt-block metadata before local lookup. Lookup selects the preceding duplicate, then asks the vendor outer manager for usable cached lengths. Release returns inner IDs in reverse allocation order. Decode also reserves the vendor dummy block. These details matter: logical token capacity alone does not determine a local hit.

Prompt hashes are chained SHA-256 digests of compact JSON token blocks. Only $\lfloor(P-1)/128\rfloor$ complete blocks are hashed for a prompt of $P$ tokens; one query token remains to compute. Generated-token blocks receive no reusable hashes. The model thus ignores content-dependent reuse after the prompt.

The planner's host directory holds
\[
 M=\left\lfloor\frac{C\,2^{30}}{128\cdot 147456}\right\rfloor
\]
slots for capacity $C$ in GiB. It stores hash presence, not payload bytes. Lookup touches resident keys in reverse prefix order and returns the longest contiguous prefix. On completion, missing keys enter in prefix order and the oldest entries are evicted as needed. An insertion larger than the whole directory is skipped. The prospective tests use $C=0$ or 36, so the directory is not validated for small capacities.

Let $H_l$ and $H_h$ be usable local and host prefix lengths. The remaining native prefill call count is
\[
 m=\left\lceil\frac{P-\max(H_l,H_h)}{128}\right\rceil.
\]
If $H_h>H_l$, restore cost covers all $H_h$ tokens, not merely $H_h-H_l$, matching the implemented transfer path. Otherwise no host load is charged. Store cost covers the prompt blocks absent from the host directory when the operation starts. The engine stays busy for the combined load, prefill and store duration; other requests on that engine cannot decode during it. Missing host entries are published only at completion. The first generated token and any request completion are then processed. The model does not expose separate intermediate load/prefill/store events, transfer overlap or concurrent host loads.

\subsection{Costs and Length Distributions}

Calibration uses the seed-701 cohort: four families $\times$ two policies (B0 and F6+36\,GiB), eight runs total. Each run has 480 measured and 32 warmup workflows. Offered rates are 0.90, 0.45, 0.55 and 0.55 workflows/s for chain, fork, mixed and mixed-burst. We pool successful observations, including warmup, without reweighting runs or fitting coefficients to response-time error.

For role $r$, phase $p$ and active count $n$, $u_{r,p,n}$ is the arithmetic mean of successful host-inclusive runtime-call durations. The wrapper residual $o_{r,p,n}$ is the mean of $\max(0,\text{model span}-\sum\text{runtime spans})$ for the corresponding model calls. A role-specific $d_r$ is the mean gap between consecutive decode model calls with a surviving request. It is an observed gap and may include more than scheduler time. The model adds it to each model operation, including prefill.

Table~\ref{tab:model-costs} gives the lookup values. The installed decode graphs use batch shapes 1, 2 and 4, but the simulator indexes observations by active count 1--4; three active requests retain their own measured coefficient. Padding and graph-selection costs are absorbed in that observation.

\begin{table}[ht]
\centering\small
\setlength{\belowcaptionskip}{6pt}
\caption{Frozen runtime mean $u$ and wrapper residual $o$, in milliseconds. Prefill $u$ is per 128-token runtime call; prefill $o$ is per model forward, not per chunk.}
\label{tab:model-costs}
\setlength{\tabcolsep}{4pt}
\begin{tabular}{@{}llrrr@{}}
\toprule
Role & Phase & Active & $u$ & $o$ \\
\midrule
Planner & prefill & 1 & 57.876607 & 1.760586 \\
Planner & decode & 1 & 33.706571 & 0.117403 \\
Planner & decode & 2 & 34.441473 & 0.118596 \\
Planner & decode & 3 & 36.734474 & 0.135935 \\
Planner & decode & 4 & 37.070009 & 0.116141 \\
Executor & prefill & 1 & 14.958987 & 1.875139 \\
Executor & decode & 1 & 5.964320 & 0.110593 \\
Executor & decode & 2 & 8.814087 & 0.113186 \\
Executor & decode & 3 & 10.011593 & 0.130984 \\
Executor & decode & 4 & 12.514483 & 0.115294 \\
\bottomrule
\end{tabular}
\end{table}

The dispatch gaps are $d_\mathrm{planner}=1.380052$\,ms and $d_\mathrm{executor}=1.087696$\,ms. Planner host-store and host-load coefficients are $s=84.355457$ and $\ell=66.538727$ microseconds/token. Each is total observed path time divided by total transferred tokens. Loads use the full available host prefix when restoration occurs. The coefficients include packing, unpacking and connector overhead as well as the copy itself.

For a prefill with $S$ missing-store tokens and $R=H_h\mathbf{1}[H_h>H_l]$ restored tokens, the model charges
\begin{align*}
 D_\mathrm{prefill}&=m u_{r,\mathrm{prefill},1}
                   +o_{r,\mathrm{prefill},1}+d_r\\
                  &\quad+R\ell+Ss,\\
 D_\mathrm{decode}(n)&=u_{r,\mathrm{decode},n}
                     +o_{r,\mathrm{decode},n}+d_r.
\end{align*}
Host terms are zero for the executor and for host-OFF runs. A snapshot of the current running set defines each decode batch, so occupancy changes only through arrivals and completions.

The \texttt{usage.completion\_tokens} values of successful responses give the length histograms, grouped by role, node name and whether prompt length exceeds 1,000 tokens. Each draw is capped at that request's generation budget. All seven calibration classes are degenerate: short-context plan, execute and answer have 32 tokens; long-context plan has 32, each fork execute branch 128, and answer 64. Each class contains 2,048 observations. The predictor therefore does not reproduce output-length variation or semantic output differences. Draws occur in event order with seed 1701 and use no held-out responses. The final model retains its 16.0\% in-sample fork-B0 P99 underestimate.

\subsection{Event-Loop Pseudocode}

The following specifies the reduced engine. Cache allocation and publication follow the rules above; all times use the coefficients just defined.

\begingroup\small
\begin{enumerate}
\setlength{\itemsep}{2pt}
\setlength{\parskip}{0pt}
\item Pop the earliest event, breaking ties by insertion number. Advance time and execute its callback.
\item On workflow arrival, enqueue FIFO. Admit while the policy's flow cap allows. For newly ready nodes, draw a length and schedule send after the tool delay.
\item On send, enqueue at the role cap if necessary; otherwise append to that engine's FIFO queue and pump it.
\item Pump: if busy, return. If a running slot is free, try allocating the oldest waiting request. On success, move it to running, compute $H_l,H_h,m,R,S$, mark busy, and schedule prefill completion after $D_\mathrm{prefill}$. Return.
\item Otherwise, if requests are running, extend their native allocations and obtain the dummy block. Capture this batch, mark busy and schedule decode completion after $D_\mathrm{decode}(n)$.
\item Prefill completion: publish missing host keys, set the request's generated count to one, and clear busy. Decode completion: clear busy and increment each captured request's count.
\item Release requests that reach their sampled length. Queue their dependency/role-slot callbacks at the same time, then pump the engine again. Completed workflows release a flow slot. Continue until all events drain.
\end{enumerate}
\endgroup

\subsection{Freezes and Finite-Window Scoring}

\paragraph{Common load.} The source, calibration JSON, cache snapshot, RNG seed, traces, 24 conditions and predictions were hashed before inference. These cover four families, three policies and seeds 702/703, each with 1,200 measured and 32 warmup workflows. Rates are 0.72 for chain, 0.36 for fork and 0.44 for mixed/burst. Per-arm latency error is $(\hat y-y)/y$, with frozen absolute tolerance 0.15; SLO tolerance is 0.05. Policy deltas are scored separately. No measured held-out length, batch occupancy, cache hit or completion order enters a prediction. The client I/O repair preceded measured data and did not change predictions; its effect relative to the calibration client is not modeled.

Predicted workflow latency is completion time minus its scheduled release, including admission wait. P50 and P99 use linear interpolation between sorted observations. Completion throughput is the measured workflow count divided by the interval from the first measured arrival to the final completion, including drain. SLO counts all measured workflows. Board energy is outside the model.

\paragraph{Endpoints.} The simulator, cache model and calibration JSON are byte-identical to those of the common-load predictor. The screen chooses endpoints; before confirmation, both endpoints of all 12 brackets, their traces, seeds 712/713, 48-run order and predictions are frozen. Runs have at least 1,200 workflows and 1,800\,s of arrivals. Predicted completion records are passed to the same judge used for measured records. Simulation release times are reconstructed as completion minus latency, and every simulated completion is marked successful. This predicts a finite-window outcome under the model, not a timeout-aware capacity limit.

For $N$ offered workflows and declared arrival window $T$, the judge requires zero request/workflow errors, at least 0.99 SLO attainment in each workflow class, and bounded backlog and drain. It samples released-minus-finished workflows at 141 evenly spaced points in $[0.2T, 0.9T]$; bursts instead use 60-second cycle boundaries in that interval. Fewer than five samples are inconclusive. With $m_q=\max(1,\lfloor n_q/7\rfloor)$, it compares the mean of the last $m_q$ backlog samples to the first $m_q$. This growth and the least-squares slope multiplied by $T$ must each be at most $\max(6, 0.01N)$. Drain after $T$ must be at most $\max(20\,\mathrm{s},0.03T)$. Invalid IDs, clocks or instrumentation are inconclusive. Missing completions after a measured hard timeout fail; the model has no such timeout mechanism. Numerical errors use $|y-\hat y|/\hat y$ with tolerance 0.15, and SLO tolerance 0.05, separately from the pass/fail call.

\paragraph{Store change.} The panel keeps the same model, lengths, cache rules, load cost and runtime coefficients. Only A0's store coefficient is multiplied by the op gate's geometric-mean ratio, 0.620109739, as declared in advance. The eight-run fork panel uses rate 0.36, seeds 704/705 and two repetitions per variant per seed. Current/A0/A0/current is followed by A0/current/current/A0. Traces and order are fixed in advance; the gate-derived predictions are hashed before the first panel run. This cohort uses the predicted-denominator numerical error above. Scaling an isolated operation's cost is a testable transfer assumption, not a measurement of its serving effect.

\paragraph{Limits.} Mean costs omit rare stalls, request failures, timeout/abort recovery and preemption. The model does not simulate tensor arithmetic, transport queues, output-dependent hashes or scheduling effects from semantic branching. The same judge cannot restore these missing mechanisms. We report numerical fit, policy direction and endpoint classification separately; none of them is a confidence interval or a proof of stationary stability.